\documentclass[twocolumn]{aastex62}
\usepackage{comment}

\newcommand{\mstar}{M$_{\star}$}

\newcommand {\msun}{M$_{\odot}$}
\newcommand{\numberofsatellites}{63}
\newcommand{\numberofhalos}{six }
\newcommand{\velocityspan}{1000 km s$^{-1}$ }
\newcommand{\densityspan}{6 dex }
\newcommand{\rpspan}{5 dex }
\newcommand{\boxname}{high-resolution forced-refined box}
\usepackage{graphics}
\usepackage{lipsum}
\graphicspath{{./}{./}}

\defcitealias{peeples19}{Paper~I}
\defcitealias{corlies18}{Paper~II}
\defcitealias{zheng20}{Paper~III}

\received{\today}
\revised{TBD}
\accepted{TBD}
\submitjournal{ApJ}
\usepackage{ragged2e}
\shorttitle{FOGGIE IV: Ram Pressure Stripping in Galactic Halos}
\shortauthors{Simons et al.}

\begin{document}

\title{Figuring Out Gas \& Galaxies In Enzo (FOGGIE). IV. The Stochasticity of Ram Pressure Stripping in Galactic Halos}

\correspondingauthor{Raymond C. Simons}
\email{rsimons@stsci.edu}

\author[0000-0002-6386-7299]{Raymond C.\ Simons}
\affil{Space Telescope Science Institute, 3700 San Martin Drive, Baltimore, MD 21218}

\author[0000-0003-1455-8788]{Molly S.\ Peeples}
\affiliation{Space Telescope Science Institute, 3700 San Martin Drive, Baltimore, MD 21218}
\affiliation{Department of Physics \& Astronomy, Johns Hopkins University, 3400 N.\ Charles Street, Baltimore, MD 21218}

\author[0000-0002-7982-412X]{Jason Tumlinson}
\affiliation{Space Telescope Science Institute, 3700 San Martin Drive, Baltimore, MD 21218}
\affiliation{Department of Physics \& Astronomy, Johns Hopkins University, 3400 N.\ Charles Street, Baltimore, MD 21218}

\author[0000-0002-2786-0348]{Brian W. O'Shea}
\affiliation{Department of Computational Mathematics, Science, and Engineering, 
Department of Physics and Astronomy, 
National Superconducting Cyclotron Laboratory,  
Michigan State University}

\author[0000-0002-6804-630X]{Britton D. Smith}
\affiliation{Royal Observatory, University of Edinburgh, United Kingdom}

\author[0000-0002-0646-1540]{Lauren Corlies}
\affiliation{Rubin Observatory Project Office, 950 N. Cherry Ave., Tucson, AZ 85719}

\author[0000-0003-1785-8022]{Cassandra Lochhaas}
\affiliation{Space Telescope Science Institute, 3700 San Martin Drive, Baltimore, MD 21218} 

\author[0000-0003-4158-5116]{Yong Zheng}
\affiliation{Department of Astronomy, University of California, Berkeley, CA 94720}
\affiliation{Miller Institute for Basic Research in Science, University of California, Berkeley, CA 94720}

\author[0000-0001-7472-3824]{Ramona Augustin}
\affiliation{Space Telescope Science Institute, 3700 San Martin Drive, Baltimore, MD 21218}

\author[0000-0001-7472-3824]{Deovrat Prasad}
\affiliation{Department of Computational Mathematics, Science, and Engineering, 
Department of Physics and Astronomy, 
National Superconducting Cyclotron Laboratory,  
Michigan State University}

\author[0000-0001-7472-3824]{Gregory F.\ Snyder}
\affiliation{Space Telescope Science Institute, 3700 San Martin Drive, Baltimore, MD 21218}

\author[0000-0002-9599-310X]{Erik Tollerud}
\affiliation{Space Telescope Science Institute, 3700 San Martin Drive, Baltimore, MD 21218}

\begin{abstract}
We study ram pressure stripping in simulated Milky Way--like halos at $z\ge 2$ from the Figuring Out Gas \& Galaxies In Enzo (FOGGIE) project. These simulations reach exquisite resolution in their circumgalactic medium (CGM) gas owing to FOGGIE's novel refinement scheme. The CGM of each halo spans a wide dynamic range in density and velocity over its volume---roughly \densityspan and 1000 km s$^{-1}$, respectively---translating into a \rpspan range in ram pressure imparted to interacting satellites. The ram pressure profiles of the simulated CGM are highly stochastic, owing to kpc-scale variations of the density and velocity fields of the CGM gas. As a result, the efficacy of ram pressure stripping depends strongly on the specific path a satellite takes through the CGM. The ram-pressure history of a single satellite is generally unpredictable and not well correlated with its approach vector with respect to the host galaxy. The cumulative impact of ram pressure on the simulated satellites is dominated by only a few short strong impulses---on average, 90\% of the total surface momentum gained through ram pressure is imparted in 20\% or less of the total orbital time. These results reveal an erratic mode of ram pressure stripping in Milky-Way like halos at high redshift---one that is not captured by a smooth spherically-averaged model of the circumgalactic medium.
\end{abstract}


\section{Introduction}\label{sec:intro}

When a galaxy passes through a gaseous medium, such as the intracluster medium of a massive cluster or the circumgalactic medium (CGM) of another galaxy, it will encounter a headwind known as ``ram pressure.'' The strength of this headwind is set by the gas density of the local medium and its relative speed with the galaxy. Both of these will vary with time---as the galaxy speeds up or slows down, as it samples other parts of the medium, and as the medium evolves.

\citet{1972ApJ...176....1G} proposed that ram pressure can remove gas from a galaxy when and where its strength exceeds the galaxy's own local gravitational restoring pressure. This action is known as ram pressure stripping and is more effective in lower mass galaxies, where gravitational restoring pressures are generally lower, and for galaxies moving through more massive host halos, where ram pressures are generally higher.

Ram pressure stripping plays a key role in regulating the gas content and star formation activity of the low mass satellite populations around galaxies like the Milky Way and their galaxy groups. It is thought to be responsible for removing gas from Local Group dwarf galaxies as they approach the Milky Way and M31 \citep{2003AJ....125.1926G, 2006MNRAS.369.1021M}, in turn shutting down their star-formation \citep{2015MNRAS.454.2039F, 2016MNRAS.463.1916F, 2018MNRAS.478..548S}. This is motivated by the fact that dwarf galaxies near the Milky Way and M31 tend to be more gas-poor than those at large distances \citep{2009ApJ...696..385G}. In certain circumstances, ram pressure can also re-ignite star-formation in previously non-star-forming dwarf galaxies \citep{2019MNRAS.482.1176W}.

The impact of ram pressure stripping on the galaxies within galaxy groups has been studied extensively with controlled numerical wind tunnel experiments (e.g., \citealt{2015ApJ...815...77S, 2016ApJ...826..148E, 2018ApJ...867...72W, 2018ApJ...863...49B, 2019A&A...624A..11H}) and semi-analytic models of galaxy formation (e.g., \citealt{2008MNRAS.383..593M, 2015MNRAS.451.2663H, 2017MNRAS.471..447S}). Such models have been used to simulate the ram pressure histories of Local Group dwarf galaxies and back out the properties of the Milky Way's circumgalactic medium (e.g., \citealt{2000ApJ...541..675B, 2013MNRAS.433.2749G, 2015ApJ...815...77S}).

With a few exceptions (e.g., \citealt{2019MNRAS.487.4313A}), however, these models treat the CGM (or intra-group medium) as a smoothly-varying and spherically-symmetric fluid with velocities and densities obeying monotonic functions. As a result, the strength of ram pressure varies with time in a smooth and predictable manner and is analytically derivable (or at least easily computable) from a satellite's trajectory through the medium.  

However, in the last decade, observations and cosmological numerical simulations have started to reveal a more complex picture of the CGM---as a rich multiphase medium with intricate small-scale structure (see \citealt*{review17} and references therein). This picture of the CGM is challenging to capture in the general form of an analytic model and raises important questions about the true nature of ram pressure stripping in such a medium. In this paper, we study ram pressure stripping of satellites using a suite of six cosmological zoom-in galaxy formation simulations of Milky Way-like halos from the Figuring Out Gas \& Galaxies in Enzo (FOGGIE) project. These simulations enforce exquisite spatial (and, as a consequence, temporal) resolution in the CGM, revealing the multiphase and multiscale nature of its gas in great detail. Over the volume of the CGM, these simulations show a wide dynamic range of densities and velocities, and as a consequence, ram pressures.

In \S\ref{sec:foggie_simulations}, we introduce the six halos of the FOGGIE simulation suite and discuss differences in the simulation setup from previous generations of FOGGIE. Then, in \S\ref{sec:RP_section}, we discuss a generalization of the criterion for ram pressure stripping and highlight the large distribution of ram pressures present in a typical FOGGIE simulated CGM. In \S\ref{sec:rp_profiles}, we examine the radial ram pressure profiles of the FOGGIE halos, using randomly sampled trajectories through the static $z\,=\,2$ simulation snapshots, and assess their impact on a population of toy satellites. Next, in \S\ref{sec:FOGGIE_satellite_galaxies}, we characterize the real FOGGIE satellite galaxy population at $z\,=\,2$ and track their orbits back through the simulation box. We assess the impact of ram pressure on the FOGGIE satellite population and compare it against the ram pressure that would be inferred using a spherically-averaged analytic model of their respective halos. Finally, in \S\ref{sec:conclusions}, we summarize the results of the paper.

\begin{deluxetable*}{lcccccccccc}[th]
\tablenum{1}
\tablecaption{FOGGIE Simulations\label{tab:halo_table}}
\tablewidth{0pt}
\tablehead{& & \multicolumn{1}{|c|}{z = 0}  &  \multicolumn{8}{c|}{z = 2}  \\ & & \multicolumn{1}{|c|}{} & &   &  & & &  \multicolumn{3}{|c|}{inside force-refined box}  \\ \colhead{Halo} & Environment & M$_{200}$   & R$_{200}$   & M$_{200}$   & \colhead{\mstar}  & \colhead{M$_{\text{ISM}}$} & \colhead{M$_{\text{CGM}}$} &  \multicolumn{2}{c}{f$_{\text{CGM, with CL resolved}}$} &  N$_{\text{sats}}$  \\ 
& &  ($10^{12}\,$ \msun)  & (pkpc)  &($10^{12}\,$ \msun)&($10^{10}\,$ \msun) &($10^{10}\,$ \msun) & ($10^{10}\,$ \msun) & (volume, \%)& (mass, \%) & }
\startdata
Hurricane & Overdense    &  1.50 & 82.2 & 0.52 & 2.75 & 0.56 & 2.82 & 99.9  & 95.9 &  20 \\
Cyclone   & Underdense   &  1.44 & 91.5 & 0.73 & 3.21 & 0.43 & 3.48 & 99.7  & 91.5 &  22 \\
Blizzard  & Underdense   &  1.34 & 77.2 & 0.48 & 4.89 & 1.06 & 1.16 & 99.9  & 96.9 &  6  \\
Squall    & Mean         &  1.09 & 55.3 & 0.16 & 0.92 & 0.55 & 0.67 & 99.9  & 98.8 &  7  \\
Maelstrom & Mean         &  1.16 & 69.9 & 0.33 & 3.25 & 0.67 & 1.10 & 99.9  & 97.6 &  9  \\
Tempest   & Mean         &  0.60 & 51.8 & 0.14 & 1.56 & 0.16 & 0.39 & 99.9  & 98.8 &  5  \\
\enddata
\tablecomments{Properties of the six FOGGIE simulations. The environment label is determined using nearest neighbor statistics, where ``mean'', ``overdense'', and ``underdense'' are at the mean, mean $+ 1\sigma$, and mean $- 1\sigma$ of all Milky-Way like halos in the 100 Mpc h$^{-1}$ pathfinder simulation, respectively. The mass of the stars (\mstar) and the interstellar medium (M$_{\text{ISM}}$) are measured in a 10 kpc sphere around the central galaxy at $z=2$, where M$_{\text{ISM}}$ only includes cold gas ($T < 1.5 \times 10^{4}$ K). The mass of the circumgalactic medium (M$_{\text{CGM}}$) is the mass of gas inside the virial radius (R$_{200}$) at $z=2$, not including the cold interstellar media of the central and satellite galaxies. The fraction of the CGM volume and gas mass inside the force refinement region where the cooling length is resolved  (see text) is listed, as well as the number of satellite galaxies detected in each halo.}
\end{deluxetable*}


\begin{figure*}
\includegraphics[width=\textwidth]{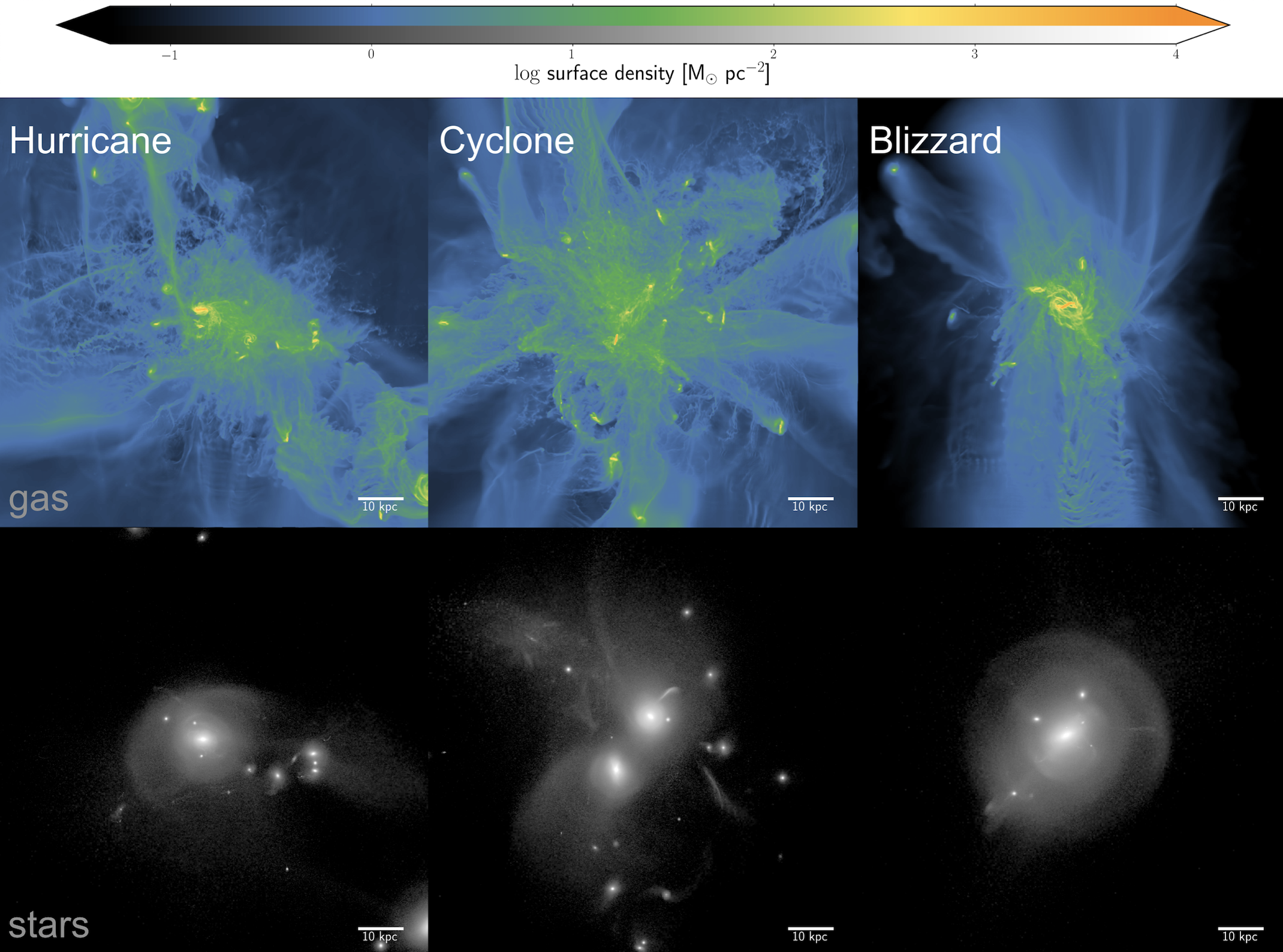}
\caption{Projections of the gas and star surface mass densities of the Hurricane, Cyclone and Blizzard FOGGIE halos at $z=2$ are shown. The box is the size of the force-refined region and the white bar marks 10 kpc. The properties of the six FOGGIE halos at $z=2$ are listed in Table \ref{tab:halo_table}.\label{fig:six_halos_projections}}
\end{figure*}

\renewcommand{\thefigure}{\arabic{figure} (cont.)}
\addtocounter{figure}{-1}

\begin{figure*}
\includegraphics[width=\textwidth]{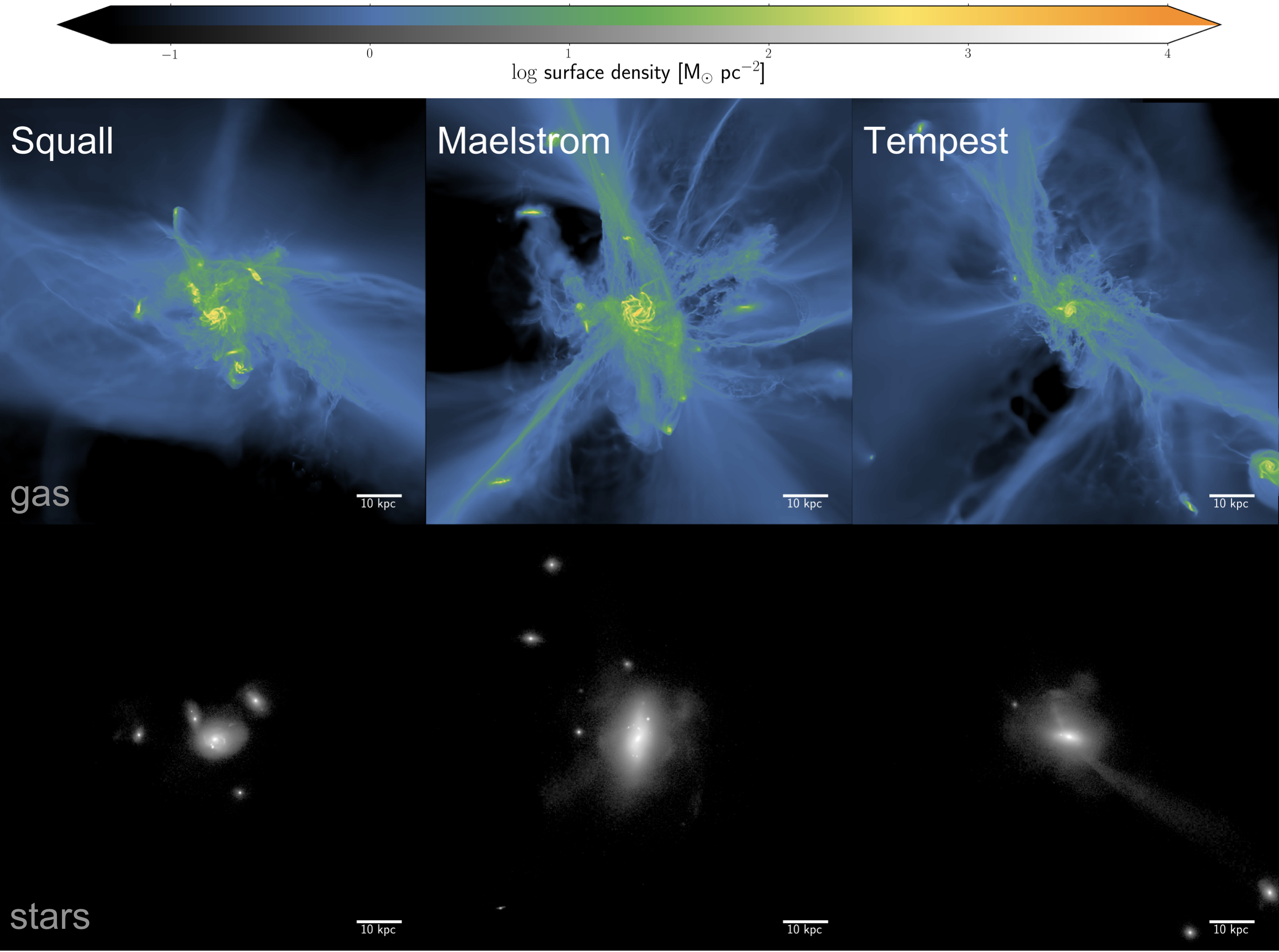}
\caption{ A continuation of Figure \ref{fig:six_halos_projections}. Projections of the gas and star surface mass densities of the Squall, Maelstrom and Tempest FOGGIE halos at $z=2$ are shown.\label{fig:six_halos_projections_cont}}
\end{figure*}

\renewcommand{\thefigure}{\arabic{figure}}

\begin{figure*}
\centering
\includegraphics[width=\textwidth]{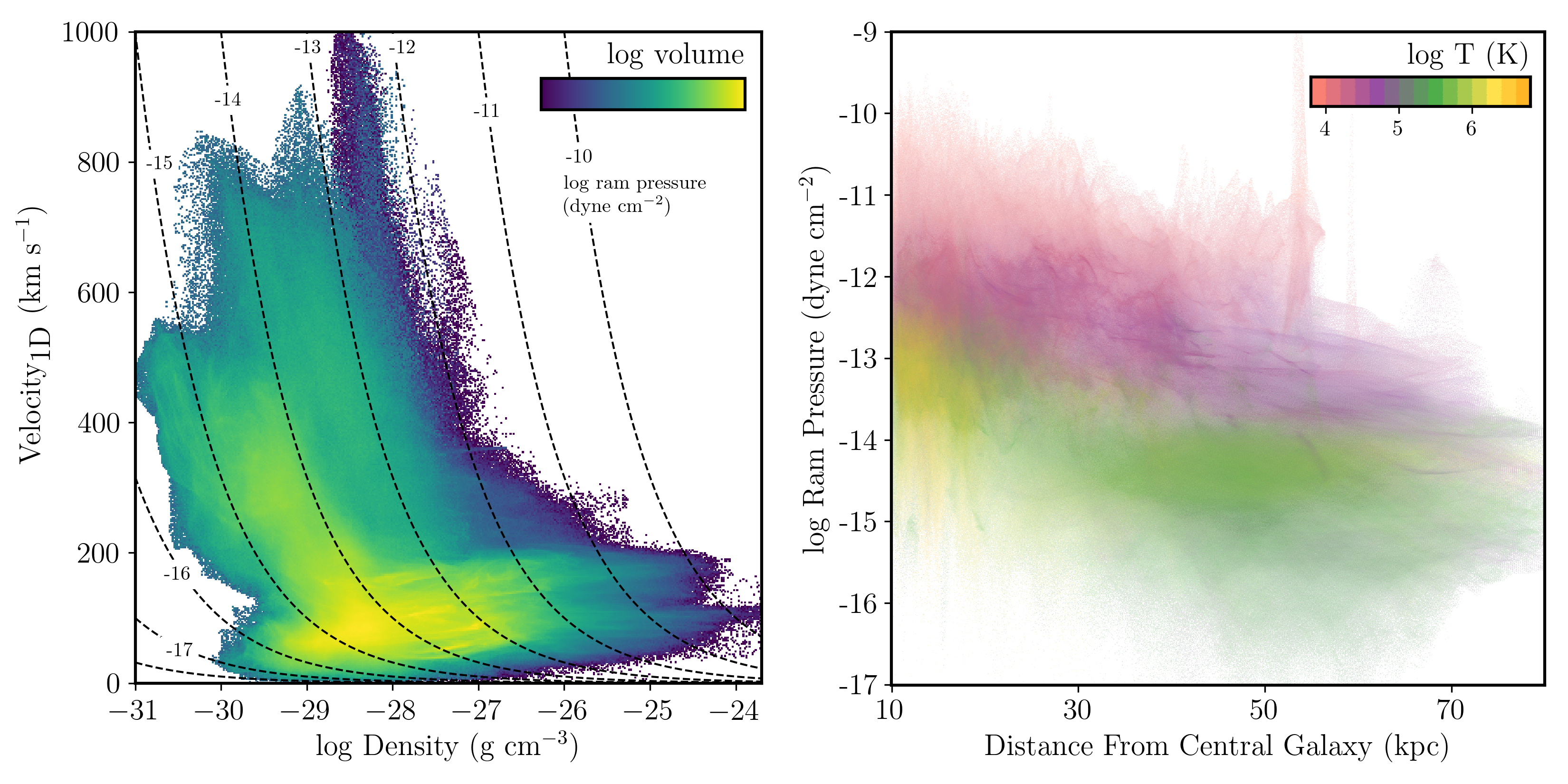}
\caption{Left panel: The volume-weighted distribution of density and 1D velocity ($|\vec v|$/$\sqrt3$) of circumgalactic gas in the Tempest halo at $z=2$. Velocities are measured in the rest-frame of the halo. Lines of constant ram pressure are included---the quoted strength is that acting on an object with zero velocity in the halo rest-frame. The FOGGIE simulations recover a large range of densities and velocities in the CGM. This translates into a roughly \rpspan range of ram pressure. Right panel: The ram pressure experienced by an object at rest in the CGM as a function of the object's distance from the central galaxy. The average temperature of the CGM gas is color-coded, with colder gas in pink and hotter gas in orange. The strength of ram pressure increases, on average, towards the halo center, with the highest ram pressure coming from the cold dense gas. At any given position, the CGM spans several orders of magnitude in ram pressure strength.}\label{fig:velocity_density_RP_CGM}
\end{figure*}

\section{FOGGIE Simulations}\label{sec:foggie_simulations}

In this section, we outline the details of the simulations used in this paper (\S\ref{sec:sim_details}, Table \ref{tab:halo_table}, Figure \ref{fig:six_halos_projections}), and compare the simulated halo and galaxy masses with those of Milky-Way like galaxies in the present-day and at $z=2$ (\S\ref{sec:mw_comparison}).

\subsection{Simulation Details}\label{sec:sim_details}
We analyze six distinct FOGGIE simulations zoomed-in on halos with $z = 0$ virial masses similar to the Milky Way. These simulations were run with the open-source adaptive mesh refinement (AMR) code Enzo \citep{bryan14, brummel19}. The first generation of FOGGIE simulations is described thoroughly in \citet{2019ApJ...873..129P}, \citet{2018arXiv181105060C}, and \citet{2020arXiv200107736Z}. We briefly review them here and note differences between the previous runs and these new runs. 

This paper is the first in the FOGGIE series to analyze all six of the first-generation FOGGIE halos, which are described in Table~\ref{tab:halo_table}. The initial conditions were derived to reach a dark-matter halo virial mass of approximately M$_{200} \simeq 0.5-1.5 \times 10^{12}$ M$_{\odot}$ at $z = 0$. The halos were selected from a single (100 Mpc/h)$^3$ domain to have undergone their last significant merger (a mass ratio of 10:1 or lower) at or before $z = 2$, reflecting the last expected major merger event of the Milky Way \citep{2018Natur.563...85H}.

The simulations are run with Enzo's native star formation prescription and supernovae thermal feedback schemes \citep{2006ApJ...650..560C}. Density and metallicity-dependent cooling and ionization rates are computed using the Grackle code \citep{smith17}, including self-shielding of gas owing to H~I opacity at $z 
\leq 15$ \citep{emerick19}. Star formation occurs in dense gas with a converging flow and gas is turned into star particles in proportion to the local gas mass, with a minimum star-particle mass of 1000\,\msun. 

FOGGIE is distinguished from other zoom-in galaxy simulations primarily by its novel AMR refinement scheme. Using Enzo's flexible AMR capability, FOGGIE forces a high level of refinement on the region near the galaxy of interest. A moving box (``track box") at fixed size and resolution follows the galaxy through the domain along a path determined from a lower-resolution track-finding run. We call this ``forced refinement.'' With forced refinement, many cells within the track box contain gas with very low densities and/or high temperatures, where the timestep and cooling time are long. This gas is over-resolved, at least as far as the cooling length (sound speed $\times$ cooling time) is concerned. To better resolve thermally unstable gas, we employ a scheme for ``cooling refinement'' which modifies the refinement criterion in the track box to place the highest resolution only in cells where the local cooling time is short compared to the sound crossing time. In practice, cells are refined such that their size is smaller than the cooling length of the gas, up to a maximum level of refinement. Thus within the track box the nominal fixed refinement and the highest level of cooling refinement are parameterized separately. We use a minimum level of refinement $n_{\rm ref}=9$ everywhere within the track box, which yields a comoving cell size of 1100 pc. Within this box, cooling refinement operates where needed up to $n_{\rm ref} = 11$ (i.e., the same as the maximum level of density refinement and thus the resolution of the interstellar medium), which gives cell sizes of 274 comoving parsec. For the simulation suite used in this paper, forced and cooling refinement are started at $z=6$. The track boxes span $-100$~h$^{-1}$ to $+100$~h$^{-1}$ comoving kiloparsecs around the halo centers.

Surface density projections of the gas and the stars for the six FOGGIE simulations are shown in Figure \ref{fig:six_halos_projections}, centered on the central galaxy. The volume-weighted distribution of the velocity and density of the CGM gas in the Tempest halo is shown in the left panel of Figure \ref{fig:velocity_density_RP_CGM}. These simulations generally recover a large dynamic range in each---roughly \densityspan in density and \velocityspan in 1D velocity for Tempest.

The detailed properties of the FOGGIE halos are listed in Table \ref{tab:halo_table}, including their virial masses at $z=0$ and several properties at $z=2$: halo virial masses, stellar and cold gas masses of the central galaxies, gas masses of the circumgalactic media, and the number of satellite galaxies detected in the track boxes. Throughout this paper, we focus on the simulation snapshots at, and prior to, $z\,=\,2$. As of this work, all of the simulations have been run (at least) to this redshift. At $z = 2$, the cooling and density refinement levels (i.e., the ISM and cool CGM) are resolved to 91 physical parsec and the forced refinement level (i.e., the warmer CGM) is resolved to 366 physical parsec. In all six halos at $z\,=\,2$, the cooling length is resolved in more than 99$\%$ and 90$\%$ of the CGM volume and mass, respectively.

\subsection{Comparison with Milky Way Mass Galaxies}\label{sec:mw_comparison}

At $z=0$, the virial masses of the FOGGIE halos span M$_{200} = 0.6-1.5 \times 10^{12}$ \msun. Estimates of the virial mass of the present-day Milky Way range from M$_{200}= 0.7-1.8 \times 10^{12}$ \msun, measured using the kinematics of its halo stars, with an average of 1.1 ($\pm$ 0.3) $\times\,10^{12}$ \msun\, (\citealt{2016ARA&A..54..529B} and references therein).

At $z=2$, the virial masses of the FOGGIE halos span M$_{200} = 0.1-0.8 \times 10^{12}$ \msun, consistent with the expected growth of halos of this mass in a $\Lambda$CDM-Planck cosmology \citep{2019MNRAS.488.3143B}. Using the average relation between halo mass and stellar mass at $z=2$ \citep{2019MNRAS.488.3143B}, the expected stellar masses of the FOGGIE centrals at $z=2$ span $0.02-0.32 \times 10^{10}$ \msun. The actual stellar masses of the FOGGIE centrals at $z=2$ are higher, spanning $0.9-4.9 \times 10^{10}$ \msun.

In summary, the virial masses of all six FOGGIE halos are consistent with virial mass estimates of the Milky Way at $z=0$, and the expected virial mass of a progenitor Milky Way halo at $z=2$. However, the stellar masses of the FOGGIE centrals at $z=2$ are higher than that expected of an average Milky Way progenitor.

\section{Ram Pressure}\label{sec:RP_section}

In this section, we first introduce ram pressure and the criterion for ram pressure stripping used in this paper (\S\ref{sec:Ram_Pressure_Stripping}). We then highlight the wide distribution of velocity, density and ram pressure in the CGM of the Tempest FOGGIE halo at $z=2$ (\S\ref{sec:ram_pressure_tempest}, Figure \ref{fig:velocity_density_RP_CGM}).

\subsection{Ram Pressure Stripping}\label{sec:Ram_Pressure_Stripping}
An object moving through a fluid of density $\rho$, with relative velocity $\Delta v$, will experience a backward-facing (i.e., opposite the direction of motion) ram pressure $P_{\rm ram}$ of
\begin{equation}
P_{\rm ram} = \rho \, (\Delta v)^2.
\end{equation}
As a satellite galaxy passes through the gaseous CGM of another galaxy, ram pressure will act to remove its gas. In the classic picture of ram pressure stripping, the gas of the satellite will be removed where ram pressure exceeds the {\emph{maximum}} gravitational restoring pressure exerted by the satellite galaxy \citep{1972ApJ...176....1G}. However, this criterion is only valid if ram pressure is applied over a sufficiently long period---at least longer than the time needed for the gas element to be pushed to the height of the maximum restoring force, attain positive total energy, and escape. This is also the time over which gas will relax back into the potential if the stripping force is removed prematurely. The length of this period is comparable to the vertical period of oscillations in the galaxy potential---a few tens of Myr in the center of a Milky Way mass galaxy and up to a few hundreds of Myr in its outskirts \citep{2018MNRAS.479.4367K}. 

However, as we will demonstrate in this paper, ram pressure in the FOGGIE halos is highly variable over short distances ($<$1 kpc) and timescales ($<$10 Myr). As such, we adopt the ``short-pulse" generalisation of the ram pressure stripping criteria outlined in \citet{2007A&A...472....5J} and \citet{2018MNRAS.479.4367K}. In general, ram pressure will impart a surface momentum density equivalent to its time integral,
\begin{equation}
{\text{Total Surface Momentum Density}} = \int P_{ram}\,dt. 
\end{equation}
\noindent In practice, we calculate the right-hand side of this equation using discrete time intervals as $\int P_{ram}\,dt  = \sum_{i} P_{ram, i}\,\Delta t$, where $\Delta t$ is the discrete timestep. The specific values of $\Delta t$ used in this paper are described in later sections, where relevant.

When the total surface momentum density exceeds the surface momentum density that a gas parcel needs to escape ($ v_{esc}\,\times\,\Sigma_{gas}$), then the parcel will be removed from the potential. This generalized criterion for ram pressure stripping is thus
\begin{equation}\label{eq:stripping_crit}
 \sum_{i = t_{init}}^{t_{final}} P_{ram, i}\,\Delta t > v_{esc}(R)\,\Sigma_{gas}(R)
\end{equation}
\noindent where $\Sigma_{gas}$ is the gas surface mass density of the ISM and v$_{esc}$ is the local escape velocity of the satellite, both of which depend on the distance $R$ from the center of the satellite galaxy.

\subsection{Ram Pressure Distribution of the Tempest Halo}\label{sec:ram_pressure_tempest}
In Figure \ref{fig:velocity_density_RP_CGM}, we show the volume-weighted distribution of densities and velocities (left panel) and the temperature-coded ram pressure profile (right panel) of the CGM gas inside the Tempest halo at $z = 2$. The velocity shown is the magnitude of the average 1D velocity (i.e., $|\vec v|$/$\sqrt3$), measured relative to the rest-frame of the halo. The strength of ram pressure shown in Figure \ref{fig:velocity_density_RP_CGM} is that acting on an object with zero velocity in the rest-frame of the halo. The CGM volume is that within the high-resolution force-refined box of the simulation, excluding gas within 10 kpc of the central galaxy and all satellite galaxies. There is no universally-accepted definition of the CGM---this definition is adopted to select against the cold interstellar gas of all galaxies inside the halo.

As shown in the left panel of Figure \ref{fig:velocity_density_RP_CGM}, the circumgalactic medium spans a large range in density and velocity---roughly \densityspan and \velocityspan, respectively. This spread translates into a wide spread in ram pressure. Lines of constant ram pressure are shown to highlight this point. The CGM of Tempest spans roughly \rpspan in ram pressure-strength. 

In the right panel of Figure \ref{fig:velocity_density_RP_CGM}, we show the 2D distribution of the potential ram pressure acting on a stationary object in the Tempest CGM, as a function of that object's radial distance from the central galaxy. This distribution is color-coded using the {\emph{most common}} temperature of the gas associated with each bin. However, we note that each point on this plot consists of gas at a range of temperatures. The breadth of ram pressures (at a fixed radial distance) is due to the inhomogeneity of the CGM, as highlighted in the left panel.

At all radii, the dynamic range of ram pressure spans several dex. The strength of ram pressure is strongly correlated with the temperature of the gas, mainly because temperature and density are locally inversely correlated in a multiphase medium that is in pressure equilibrium. The cold, dense gas contributes the strongest ram pressure, while the diffuse hot material contributes the weakest. The ram pressure distribution gradually rises towards the halo center, due to the general rise in the gas density and velocity of the CGM gas towards the center of Tempest. However, the wide range of ram pressures at a fixed radial distance means that {\emph{there is only a weak association between the radial distance of an object in the Tempest halo and the ram pressure acting on that object.}} This statement is true for all six FOGGIE halos at $z\gtrsim2$.

\section{Ram Pressure Profiles of the FOGGIE Halos at \lowercase{${z}=2$}}\label{sec:rp_profiles}

In this section, we examine the distribution of potential ram pressure histories for an arbitrary object moving through the CGM of the FOGGIE halos.

In \S\ref{sec:setting_test_particles}, we sample a large number of radial trajectories through the static $z=2$ snapshots of the FOGGIE halos and simulate the ram pressure histories of test particles following these trajectories in free-fall motion. In \S\ref{sec:sec42}, we examine the ram pressure histories of four example test particles in a single FOGGIE halo, and then expand to include all test particles in all six halos. Finally, in \S\ref{sec:43}, we compare the distribution of surface momenta imparted on the test particles through ram pressure with that needed to strip gas from satellite galaxies.

\begin{figure*}
\includegraphics[width=\textwidth]{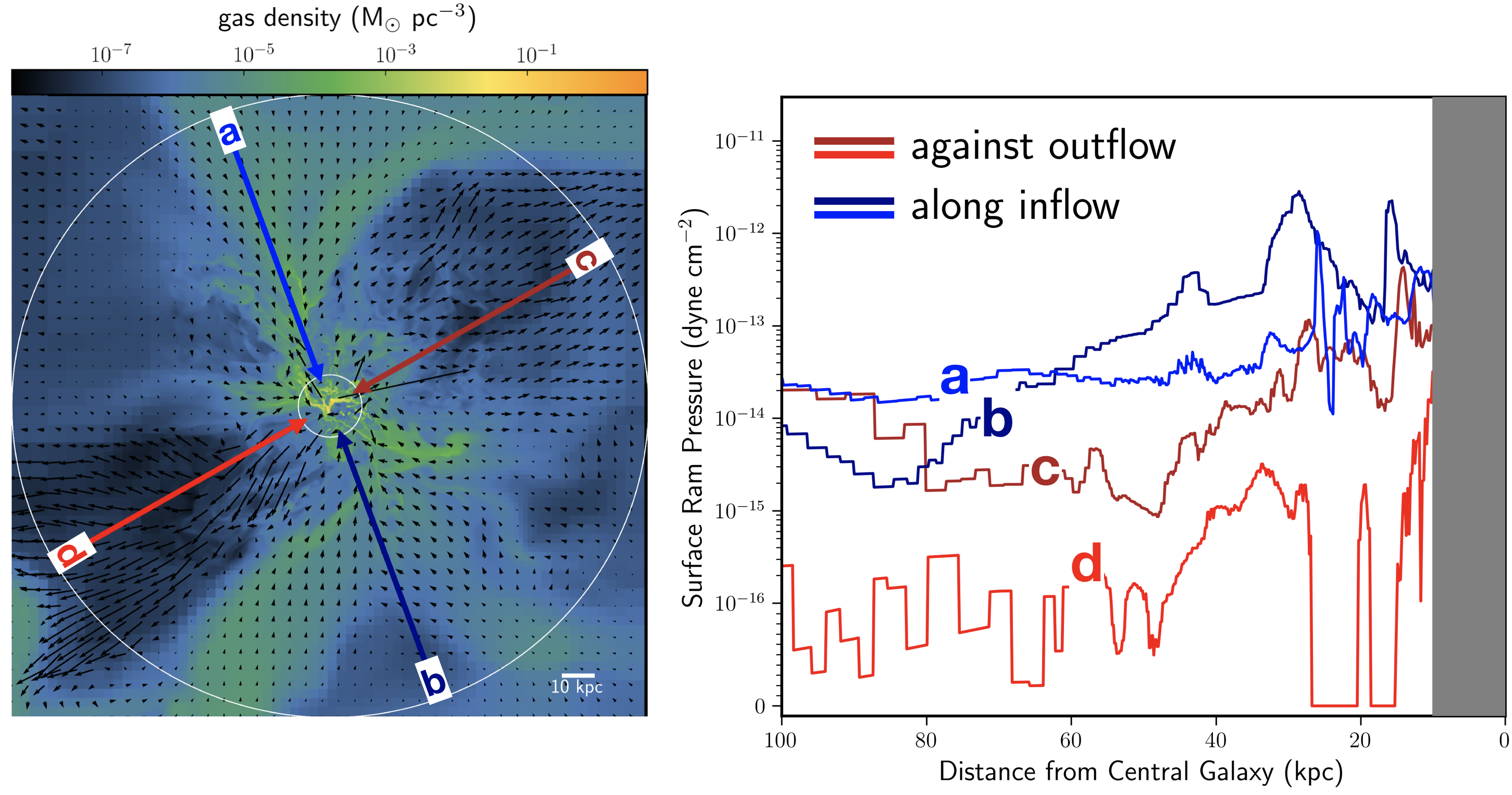}
\caption{Ram pressure profiles sampled along four trajectories through the Tempest halo. Each trajectory has a pair oriented on the opposite side of the halo.  Left panel: A slice of gas density through the simulation domain, in the plane of the trajectories, is shown. Arrows are included to highlight gas flows. The direction and size of the arrow corresponds to the velocity of the gas at that position. The large white circle marks the beginning of the trajectories, at R = 100 kpc, and the small white circle marks the end of the trajectories, at R = 10 kpc. One pair cuts through, and moves with, inflowing CGM gas (blue and dark blue lines; labeled `a' and `b') and the other pair cuts through, and moves against, outflowing CGM gas (red and dark red lines; labeled `c' and `d'). Right Panel: The ram pressure profiles of the four trajectories are shown. These profiles are complex---in addition to the significant differences between the trajectories, each shows high variability over short kpc-scale distances. On average, the ram pressure profiles are higher for the trajectories traveling along the inflowing material than for the trajectories traveling against the outflowing material. This is due to the higher average density of the inflowing material, in spite of the lower relative velocities. In general, the strength of ram pressure increases in the inner parts of the halo. The grey shaded region at $<$10 kpc is masked to avoid contamination by the interstellar medium of the central galaxy.\label{fig:RP_4traj}}
\end{figure*}

\begin{figure}
\includegraphics[width=\columnwidth]{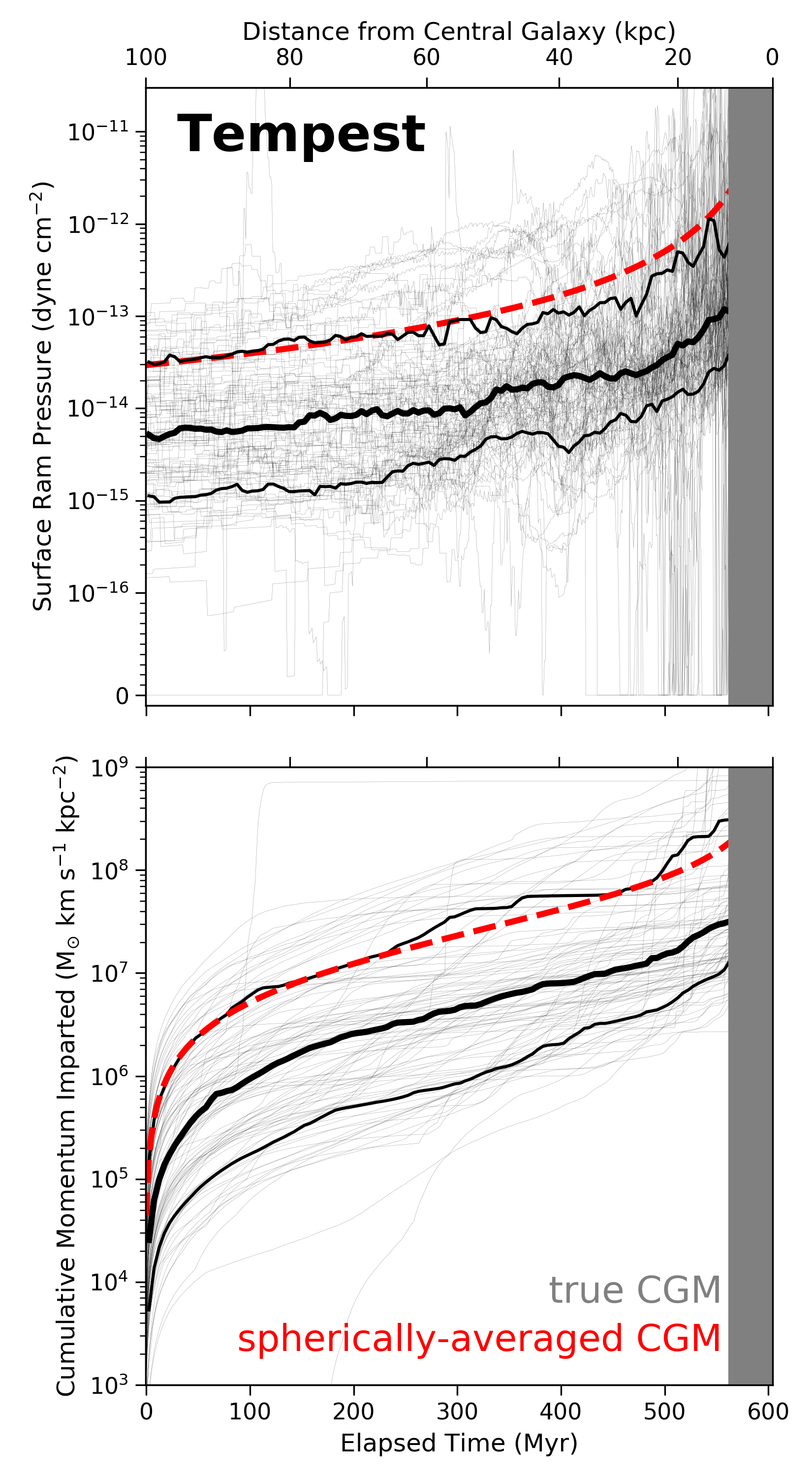}
\caption{Top: Ram pressure profiles sampled by 100 random radial trajectories through the Tempest halo at $z=2$ are shown with grey lines. The dual horizontal axis shows the elapsed time of test particles free-falling through the halo (bottom x-axis) and their distance from the central galaxy (top x-axis). The test particles are set in motion at a radial distance of 100 kpc from the central galaxy and stopped at a radial distance of 10 kpc---the latter marking the boundary between the interstellar medium of the central and its CGM (grey shaded region). The running median and 16$^{\rm {th}}$/84$^{\rm {th}}$ percentiles around the median are shown with thick and thin black lines, respectively. The equivalent trajectory through a spherically-averaged model of the Tempest CGM is shown with the red line. Bottom: The cumulative momentum surface density imparted on the test particles by ram pressure is shown for the 100 trajectories. The scatter due to trajectory-by-trajectory variations is $\sim1$\,dex. The spherically-averaged CGM generally overpredicts the accumulated momentum. \label{fig:RP_manytraj}}
\end{figure}

To summarize the following subsections, we find high variability in the ram pressure profiles sampled by individual trajectories (Figure \ref{fig:RP_4traj}) and significant $\sim$1--2 dex differences in the average ram pressure strength between trajectories (Figure \ref{fig:RP_manytraj}). There is no single, smooth ram pressure profile that can appropriately generalize the individual trajectories and a spherically-averaged model of the FOGGIE halos generally {\emph{overpredicts}} the strength of ram pressure. We compare the cumulative effect of ram pressure with toy models of satellites and find that the high trajectory-by-trajectory differences translate into large practical differences in stripping efficacy (Figure \ref{fig:total_momentum_versus_mass}).

\subsection{Setting Test Particles On Radial Free-Fall Trajectories Through the Halo}\label{sec:setting_test_particles}

To sample the ram pressure profiles of the simulations, we simulate test particles traveling on radial (i.e., towards the central galaxy) free fall trajectories through the circumgalactic medium of each halo. The test particles are set in motion at a radial distance of 100 kpc from the central galaxy and stopped at a radial distance of 10 kpc. The latter is chosen to exclude interactions with its interstellar medium. 

The velocity of the particle $V(r)$ follows
\begin{equation}
    V(r) = \sqrt{\frac{2 G M(<r)}{r}}
\label{eq:freefall}
\end{equation}
\noindent where $r$ is the distance of the particle from the central galaxy, $M(<r)$ is the total mass enclosed inside a sphere at that distance, and $G$ is the gravitational constant. Equation \ref{eq:freefall} is a lower-bound approximation of the local escape velocity and assumes spherical symmetry.

For each timestep of the simulated free fall, we measure the velocity and density of the forward-facing (i.e., in front of the direction of motion of the test particle) circumgalactic medium in the rest-frame of the test particle and record its effective ram pressure. As discussed in \S\ref{sec:RP_section}, we perform a discrete integration of the momentum surface density imparted through ram pressure. We use a relatively short timestep of 1 Myr for the integration. This ensures that the fastest-moving test particles ($\sim$500 km s$^{-1}$) travel no further than 1 grid cell between timesteps---so that they sample the simulated CGM at its full resolution.

The major advantage of this approach is its speed and flexibility. By using synthetic trajectories, we are not bound to the limited number of orbits of the real FOGGIE satellite population---allowing for a more complete sampling of the distribution of ram pressure profiles through each halo.

There are two important points to note on these test particle experiments. First, as they are run using a single simulation snapshot, they probe the {\emph{static}} ram pressure profiles of the halos. The variability of the profiles is solely due to spatial variations of the circumgalactic medium and does not include the temporal variability of the simulation. Second, these trajectories are not meant to represent realistic satellite orbits, which generally include tangential velocity components and multiple passages through the halo. By construction, the simulated trajectories represent the {\emph{minimum}} possible time an object will spend in the CGM. As a consequence, they tally the minimum average cumulative effect of ram pressure. 

Finally, we perform the same experiment using a spherically-symmetric approximation of the halo. We measure the spherically-averaged radial density profile of each $z=2$ FOGGIE halo. We use these to create a 1D model of the CGM, setting the velocity of the CGM to zero everywhere to mimic hydrostatic equilibrium. A test particle is set on a radial free fall through that model---using the velocity condition of Eq. \ref{eq:freefall}---and the effective ram pressure and accumulated momentum is recorded along the trajectory.

\subsection{Radial Ram Pressure Profiles and Cumulative Momentum Imparted}\label{sec:sec42}

\subsubsection{Individual Trajectories}\label{sec:sec421} 

In the left panel of Figure \ref{fig:RP_4traj}, we show a gas density slice through the Tempest halo at $z=2$ along with four radial trajectories in the plane of the slice. Flow lines are included to illuminate the velocity structure of the circumgalactic medium, in the rest-frame of the central galaxy. In this plane, the Tempest CGM can be coarsely segmented into regions of mainly inflow and regions of mainly outflow (both with respect to the central galaxy). We note that, while this is a convenient slice to examine the contrast in ram pressure between inflowing and outflowing gas, it is not generally true that all CGM gas is associated with a coherent flow.  The four trajectories include two that cut through (and travel in the same direction as) the inflowing CGM, and two that cut through (and travel in the opposite direction as) the outflowing CGM. These trajectories are illustrative. They are chosen by eye and do not necessarily follow the strongest or fastest parts of the flows.

In the right panel of Figure \ref{fig:RP_4traj}, we show the radial ram pressure profiles of these four trajectories. The trajectories cutting through inflowing gas show higher average ram pressure. At face value, this is counterintuitive. The relative velocities of the test particle and the circumgalactic medium are larger for those trajectories traveling against the outflowing material than it is for those traveling along with the inflowing material. However, the densities of the cold inflowing gas tend to be higher than the warm/hot outflowing gas. In this case, the density differences are more important---the ram pressure felt along the dense, cool inflowing material is higher by an average of 1--2 dex.

The trajectories have qualitative similarities. All four profiles are generally higher in the inner parts of the CGM than they are in the outer parts. This is due in part to the radial dependence on the velocity and density composition of the CGM---the static ram pressure gradually rises towards the center of the halo (Figure \ref{fig:velocity_density_RP_CGM}). It is also due to the fact that the test particles are, by construction, moving more slowly in the outer parts of the halos. The ram pressure profiles also all exhibit short-distance variability---reflecting the kpc-scale variations of the density and velocity of the Tempest CGM.  

However, the four ram pressure profiles are more dissimilar than they are similar. Specifically, they have dramatic differences in their average strength, reflecting the asymmetry of the Tempest CGM. This is the case for the paired trajectories too---those cutting through regions of the CGM with similar bulk flows. At the same radial distance in the halo, the trajectories can differ by up to 2 dex. Each of the ram pressure profiles shown is unique and highly stochastic. There is no simple generalization of their form.

To quantify the trajectory-by-trajectory scatter highlighted in Figure \ref{fig:RP_4traj}, we now examine a large number of trajectories. 

\subsubsection{Population Statistics On A Large Number of Trajectories}\label{sec:sec422} 

In the top panel of Figure \ref{fig:RP_manytraj}, we sample 100 random radial trajectories through the circumgalactic medium of the Tempest halo and simulate a test particle in free-fall along each. We show the ram pressure histories for each individual trajectory (thin grey lines), along with the running median (thick black line) and 16$^{\rm {th}}$/84$^{\rm {th}}$ percentiles (thin black lines) of the distribution of trajectories. We also include the ram pressure profile of a test particle traveling in radial free-fall through a spherically-averaged model of the Tempest CGM (red line).

As illustrated above, the individual trajectories through the true Tempest CGM are highly stochastic---showing several-dex variation in ram pressure strength over short timescales. As a result, the scatter of the population of trajectories spans $\sim$1-2 dex in ram pressure strength at a given radial distance. 

The spherically-averaged model tends to produce a ram pressure strength that is higher than the median of the distribution of true trajectories. In Tempest, specifically, the trajectory through the spherically-averaged model lies near the 84$^{\rm th}$ percentile of the distribution at all times. This result is an unavoidable consequence of the construction of the model and the manner in which the densities of the simulated CGM are distributed. Most of the mass of the Tempest CGM is concentrated in cold dense filaments and clouds. However, such high densities comprise a small fraction of the total volume of the CGM. The spherically-averaged density profile is calculated as the total mass divided by the total volume in thin successive spherical shells around the halo center. This calculation recovers the correct {\emph{total mass}} of the CGM, but is skewed high of the volume-weighted density of the CGM (i.e., the average density of a random volume element of the CGM). As a result, the inferred ram pressure of the spherically-averaged model tends to be higher than the median of the trajectories through the true CGM---the latter of which samples the volume-weighted distribution of densities.

In the bottom panel of Figure \ref{fig:RP_manytraj}, we show the cumulative momentum surface density imparted onto the test particles by ram pressure. The distribution of final surface momentum spans a little more than a dex. For the median profile of the distribution (thick black line), more than 90$\%$ of the momentum is gained in the last half of the free fall---due to the higher relative velocities and denser gas of the inner parts of the CGM. Relative to the typical trajectory through the true Tempest CGM, the test particle traveling through the spherically-averaged model (red line) accumulates a higher amount of surface momentum---an order of magnitude higher than the median profile in Tempest. These results from Tempest generalize to all six FOGGIE halos at $z=2$.

\begin{figure*}
\includegraphics[width = \textwidth]{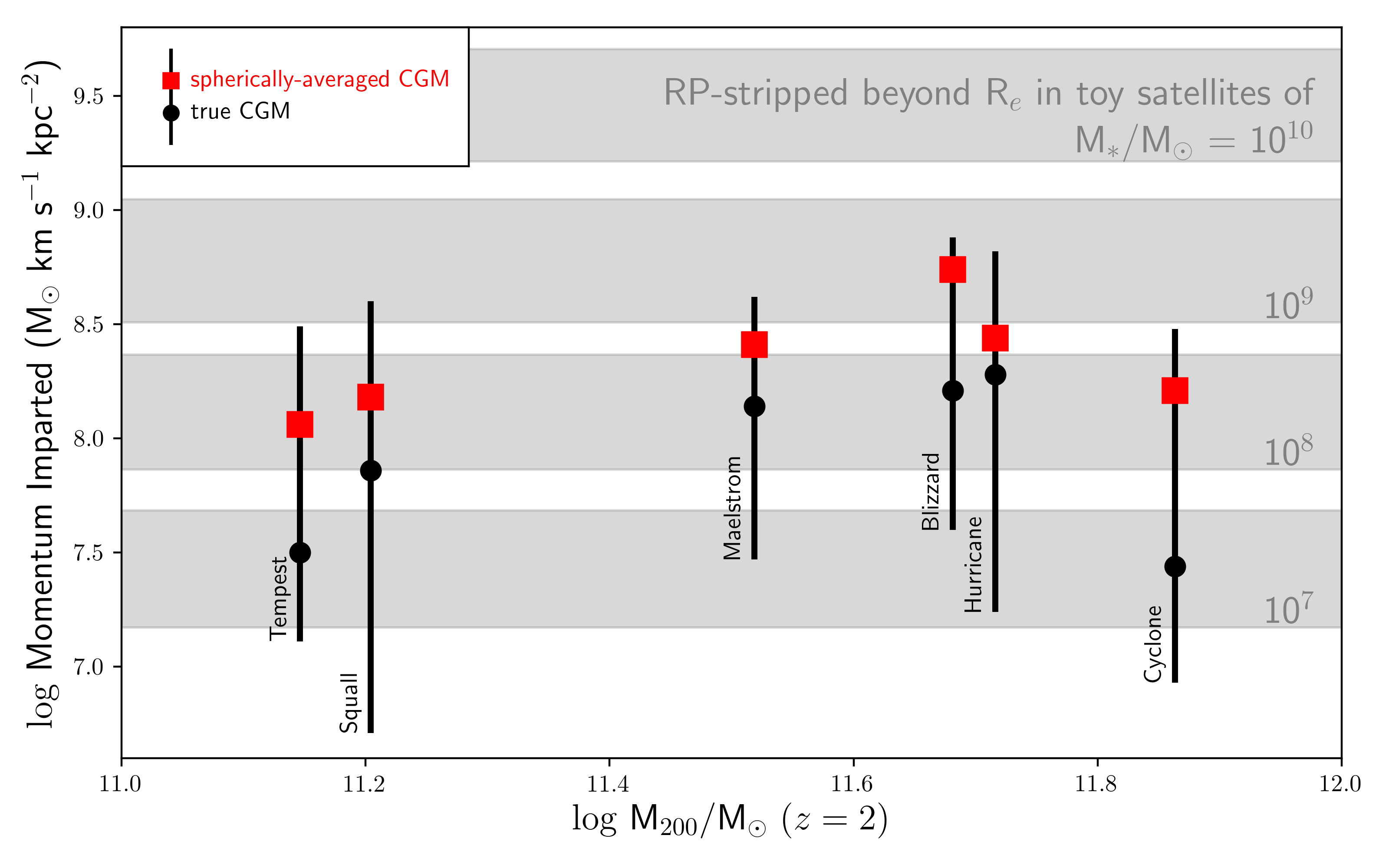}
\caption{The cumulative surface momentum imparted by ram pressure for radial trajectories through each of the six halos is shown. The black points and error bars show the median and 16$^{\rm {th}}$/84$^{\rm {th}}$ percentiles, respectively, of 100 randomly-sampled radial free fall trajectories. The red square shows a free fall trajectory through a spherically-averaged model of the CGM. The virial mass (M$_{200}$) is of the main halo at $z=2$. The surface momentum density needed to remove the interstellar medium beyond 1 effective radius (R$_{e}$) for toy satellite galaxies of masses 10$^7$ - 10$^{10}$ \msun, separated by 1 dex, are shown as grey swaths. The bottom and top of the swath show the 16$^{\rm {th}}$ and 84$^{\rm {th}}$ percentiles of the distribution of toy models at that stellar mass. For a given halo, the scatter introduced by trajectory-by-trajectory differences lead to $\sim1.5$ dex scatter in total surface momentum imparted. This scatter translates into a significant range of toy satellite masses that are susceptible to ram pressure stripping.
\label{fig:total_momentum_versus_mass}}
\end{figure*}

In Figure \ref{fig:total_momentum_versus_mass}, we show the median (black circle) and 16$^{\rm {th}}$/84$^{\rm {th}}$ percentiles (black error bars) of the final cumulative momentum of the test particles for all six FOGGIE halos at $z=2$. The mean surface momentum imparted is independent of the virial mass (M$_{200}$) of the main halo. However, we note the small dynamic range in mass probed by these halos. The scatter due to trajectory-by-trajectory variation within each halo is large and similar for all masses, spanning 1--1.5 dex. The free-fall trajectory through the spherically-averaged model of the FOGGIE halos is shown with a red square. As in Tempest, the spherically-averaged model tends to overpredict the integrated impact of ram pressure in all six halos.

\subsection{Toy Satellites}\label{sec:43}

In this subsection, we compare the distribution of the total momentum imparted from the radial trajectories (Figure \ref{fig:total_momentum_versus_mass}) with that needed to remove gas from a collection of simple toy satellite potentials. 

In \S\ref{sec:construct_toy_population}, we construct the toy satellite potentials and in \S\ref{sec:Ram_Pressure Stripping Toy_Satellites} and Figure \ref{fig:total_momentum_versus_mass}, we compare the efficacy of ram pressure stripping of these toy satellites in the FOGGIE halos.

\subsubsection{Constructing the Toy Satellite Population}\label{sec:construct_toy_population}

For a given stellar mass, we construct a population of satellites by sampling a semi-empirical distribution of their physical properties. 
We enforce a 2D single S\`ersic model \citep{1968adga.book.....S} for the distribution of baryonic mass in the satellites, following
\begin{equation}\label{eq:toy_mass_profile}
\Sigma = \Sigma_0 \exp \left(-(R/\alpha)^{1/n}\right),
\end{equation}
where $R$ is the distance from the center of the satellite, $\alpha$ is the scale length over which the mass surface density drops by a factor $e$, $\Sigma_0$ is the central surface mass density, and $n$ is the S\`ersic index which governs the shape of the distribution. 
We give half of the satellites exponential mass profiles (S\`ersic $n=1$) and half de Vaucouleurs' ($n = 4$; \citealt{1948AnAp...11..247D}). These choices bound the expected mass distributions of real galaxies.

The effective radii of the satellites (R$_{e}$, i.e., the 2D radius containing half of the total mass) are drawn from the $z=0$ size-mass relation for Sd-Irr galaxies \citep{2016MNRAS.462.1470L}, extrapolating the observed relation from its stellar mass completeness limit (at 10$^8$ \msun) to 10$^7$ \msun. For simplicity, we assume that the stellar mass effective radius is equal to the gas mass effective radius. On each draw, we include appropriate Gaussian noise to account for the observed scatter around the relation. As a note, we use the observed relation at $z=0$ instead of that at $z=2$, because the latter is not probed below $\sim10^{9.5}$ M$_{\odot}$\citep{2014ApJ...788...28V}. As such, we neglect the evolution of galaxy sizes with time. At fixed mass, high-mass late type galaxies ($\sim$10$^{10}$ M$_{\odot}$) are observed to be $\sim$3 times smaller \citep{2014ApJ...788...28V}, with slightly elevated circular velocities \citep{2016ApJ...830...14S, 2017ApJ...843...46S}, at $z=2$ than they are today. If we assume that galaxies of all masses at $z=2$ are 3 times smaller than that given from the $z=0$ relation, then the escape velocity of the toy models, and the total momentum needed for ram pressure to effectively strip them of their gas, are both underestimated in the calculations below---each by $\sim$0.2 dex. These uncertainties do not affect the conclusions of this section.

We assume a ratio of stellar mass to total baryonic mass of 0.5 inside an effective radius---generally consistent with the high mass star-forming galaxy population at $z=2$ \citep{2018ApJ...853..179T}. 

Finally, to calculate the internal gravitational force of the toy satellite, which acts to anchor gas inside the galaxy, we need to choose a ratio of baryonic mass to total mass. High mass galaxies (\mstar $\,\ge$ 10$^{10}$ \msun) at $z=2$ are thought to be strongly baryon-dominated with M$_{bary}$/M$_{tot}$ $\ge$ 0.9 in their inner parts (e.g., \citealt{2016ApJ...819...80P, 2016ApJ...831..149W}), but the same is not known for the low mass galaxy population at these redshifts (i.e., galaxies with stellar masses of $10^{7}$--10$^{10}$ \msun). The stellar mass-halo mass relation indicates that they should have lower {\emph{integrated}} baryonic-to-total mass ratios \citep{2019MNRAS.488.3143B} and they are known to be dark-matter dominated in their inner parts in the local universe \citep{2012ApJ...752...45T, 2017ARA&A..55..343B}. With these unknowns noted, we elect for a fixed baryonic-to-total mass ratio inside 1 effective radius of 0.5 across all masses.

We consider satellite galaxies that are less massive than the typical host galaxy stellar masses of the FOGGIE centrals at $z=2$ and adopt satellite stellar masses of 10$^{7}$, 10$^{8}$, 10$^{9}$, and 10$^{10}$ \msun. 

For a parcel of gas at a distance of $R$ from the center of our toy satellites, ram pressure must meet the condition laid out in Eq. \ref{eq:stripping_crit} to remove it from the satellite. We calculate the escape velocity as $v_{esc}$ = $\left(2GM(<R)/R\right)^{1/2}$, where $R$ is the distance from the center of the satellite. This calculation assumes that the mass is distributed with spherical symmetry. It ignores that it is marginally easier for gas to escape out of the side of the satellite that is closer to the central galaxy than the side that is further away.

\subsubsection{Ram Pressure Stripping In Toy Satellites}\label{sec:Ram_Pressure Stripping Toy_Satellites}

In Figure \ref{fig:total_momentum_versus_mass}, we show the total surface momentum needed to fully remove the interstellar medium beyond an effective radius in our toy satellite galaxies. For a given satellite mass, the amount of momentum that is needed varies from toy model to toy model (as they are each created from a unique draw of a distribution of physical properties). To illustrate the distribution of models, we show a grey shaded swath. The bottom and top of the swath represent the 16$^{\rm {th}}$ and 84$^{\rm {th}}$ percentiles of toy models at that mass. The median (black circle) and 16$^{\rm {th}}$ and 84$^{\rm {th}}$ percentiles (black error bars) of the free-fall test particle experiments are also shown. The red square represents the free-fall trajectory through a spherically-averaged model of the CGM of the halos.

Figure \ref{fig:total_momentum_versus_mass} represents a probabilistic process. If a random draw from the toy satellite population (grey shade) lands below a random draw from the distribution of trajectories (black error bar), the toy satellite will be stripped beyond one effective radius. The former corresponds to a potential physical model of a satellite and the latter corresponds to a potential path it would take through the CGM.
The wide range of accumulated momentum of the distribution of trajectories translates into a wide range of satellite galaxies that are susceptible to stripping.

For the lowest mass halo, Tempest, the scatter of the free fall simulations fully spans the distributions of toy models for satellite masses of $10^7$ to $10^8$ M$_{\odot}$. This indicates that a $10^8$ M$_{\odot}$ satellite might have its outer gas removed through ram pressure stripping while a $10^7$ M$_{\odot}$ satellite might not------depending on the specific path of each through the CGM of Tempest. 

The free fall trajectory through the spherically-averaged model of the Tempest CGM (red square) accumulates roughly an order of magnitude more momentum through ram pressure than the median of the trajectories through the true Tempest CGM (black circle). At face-value, the spherically-averaged model indicates that nearly all galaxies at and below a stellar mass of 10$^8$ M$_{\odot}$ would be stripped of gas beyond an effective radius. In reality, however, this occurs in a small fraction ($<25\%$) of the true trajectories.

In the remaining five halos (Squall, Maelstrom, Blizzard, Hurricane Cyclone), similar conclusions hold. The free fall trajectories of these halos typically span 1-1.5 dex in momentum imparted, corresponding to 1-2 dex in satellite masses that are susceptible to ram pressure stripping. As in Tempest, {\em the spherically-averaged model of the CGM generally overestimates the true momentum imparted through ram pressure}.

These results indicate that, {\emph{in a single halo}}, the specific path that a satellite takes through the CGM is as, or more, important in determining the impact of ram pressure as the halo mass.

\section{FOGGIE Satellite Galaxies}\label{sec:FOGGIE_satellite_galaxies}

In this section, we assess the nature and impact of ram pressure on the actual FOGGIE satellite population. We follow each satellite from the time it enters the \boxname\ of the simulation, at $z>2$, to the last snapshot we consider, at $z=2$. 

In \S\ref{sec:select_track_satellites}, we describe the method by which we select, track, and characterize the high-redshift satellite galaxy populations in the \numberofhalos FOGGIE halos. In \S\ref{sec:rp_impact_satellites}, we examine the physical properties of the satellites and the cumulative impact of ram pressure on the evolution of their gas masses. Finally, in \S\ref{sec:rapid_accumulation}, we assess the rapidity through which ram pressure acts and surface momentum is accumulated in the satellite galaxies, and compare that result against one obtained with a spherically-averaged model of the CGM.

To summarize the key result of this section, the satellites encounter highly stochastic ram pressure inside the box, supporting the results of the test particle experiments in \S\ref{sec:rp_profiles}. Typically, a series of a few short, strong impulses account for the majority of the integrated impact of ram pressure. This behavior is {\em not} captured if we consider the same satellites, on the same orbits, in a spherically-averaged model of their halos---as is traditionally used in analytic models of ram pressure stripping.

\subsection{Selecting and Tracking FOGGIE Satellites}\label{sec:select_track_satellites}

\subsubsection{Selecting Satellites at $z=2$}
Satellites appear as discernible peaks in the density distribution of the simulation box. For a static simulation snapshot, one may use a combination of gas, stars, and/or dark matter to identify these peaks. Our goal is to measure the evolution of individual satellites, which requires an association of satellites from snapshot to snapshot. As these simulations are Eulerian in nature, we are unable to track unique parcels of gas in time. Instead, we must use the simulation particles---stars and dark matter. Simulation particles have unique, immutable identities. Moreover, as they are not coupled to hydrodynamic forces of the circumgalactic medium through which they are moving, they better reflect the ballistic nature of their galaxy's orbit.  We elect to identify and track satellite galaxies using the more highly-resolved star particles. By this construction, our satellite galaxy population consists of subhalos that have formed a sufficient mass of stars ($\gtrsim 10^{5}$ \msun) by $z=2$.

To select satellites at $z=2$ of a given simulation, we adopt a custom peak-finding algorithm. We first deposit the star particles of the simulation refine box onto a fixed-resolution grid and mask cells that are lower than a threshold 3D density. We project the resulting grid along the three Cartesian directions of the simulation box, and use {\tt{astropy}}'s {\tt{photutils}} \citep{larry_bradley_2019_3568287} package to create a segmentation map for each of the three projections. For each 2D object the segmentation procedure detects, we identify the collection of star particles that are associated with it and measure their 3D center of mass. We perform this procedure twice---once using a fine sampling of the grid and a small threshold density (to detect diffuse objects) and once with a coarse sampling of the grid and a large threshold density (to detect compact objects). For diffuse objects, we adopt a cell sampling of (0.5 kpc)$^3$ and a density threshold of 0.5$\times$10$^6$ M$_{\odot}$ kpc$^{-3}$. For compact objects, we adopt a cell sampling of (1.0 kpc)$^3$ and a density threshold of 1.0$\times$10$^6$ M$_{\odot}$ kpc$^{-3}$. These values are chosen through iteration---adjusting them until the algorithm detects all satellites that are identified by-eye in the projection plots. We collate the results and create a single reduced catalog of unique satellites, matching those that are within 1 kpc of each other. 

We detect \numberofsatellites~satellites over all \numberofhalos halos at $z=2$, down to stellar masses of $\sim5\times10^6$ \msun~and separations of $\sim$1 kpc. The number of satellites detected for each halo is listed in Table \ref{tab:halo_table}.

\begin{figure}
\includegraphics[width=\columnwidth]{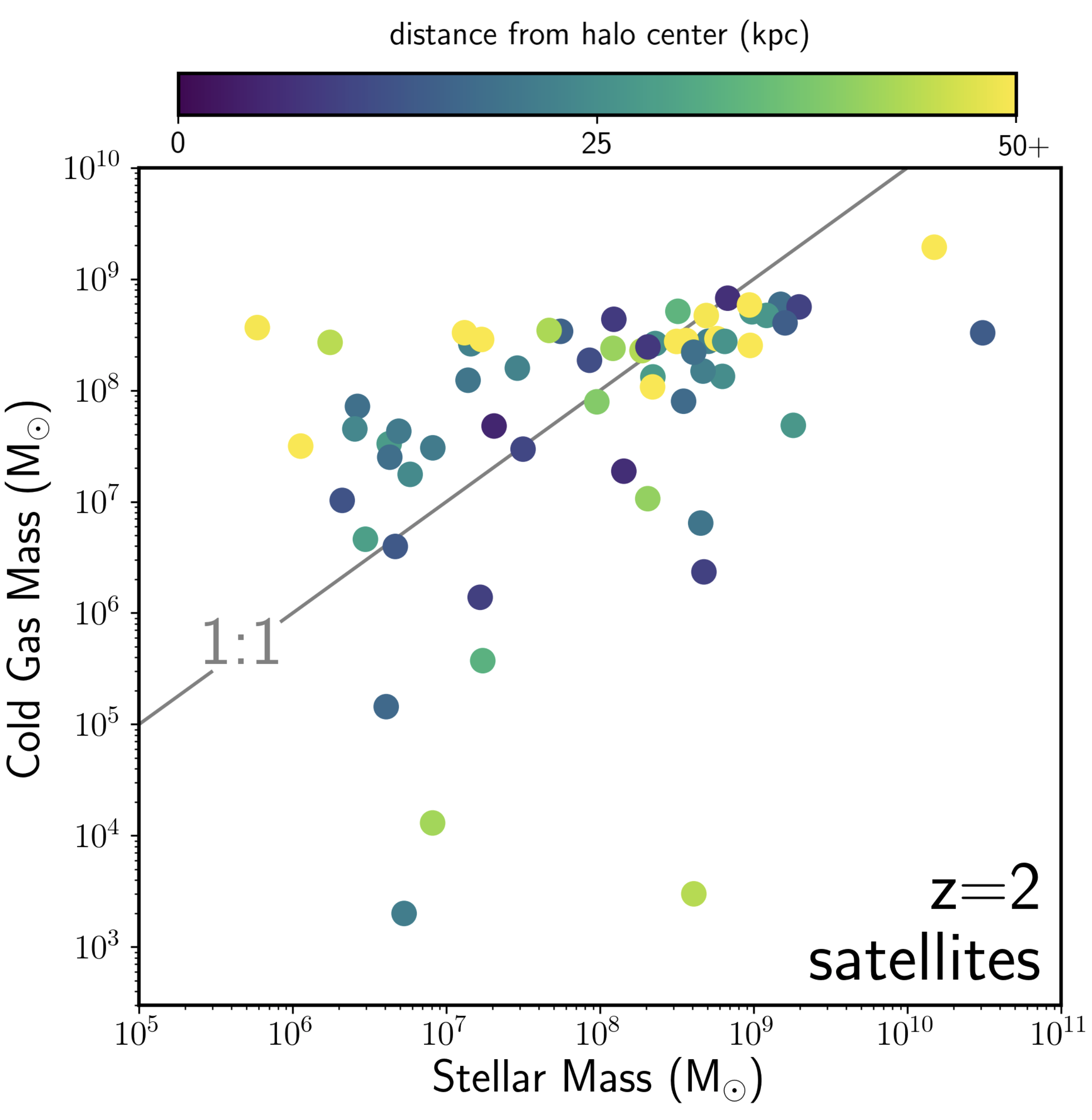}
\caption{The cold gas mass versus stellar mass for the FOGGIE satellite population, including all six halos, at $z=2$ is shown. The satellites are color-coded by the distance from their central galaxy. The 1:1 line is marked in grey. \label{fig:cold_stellar_mass}}
\end{figure}

\begin{figure*}
\includegraphics[width=\textwidth]{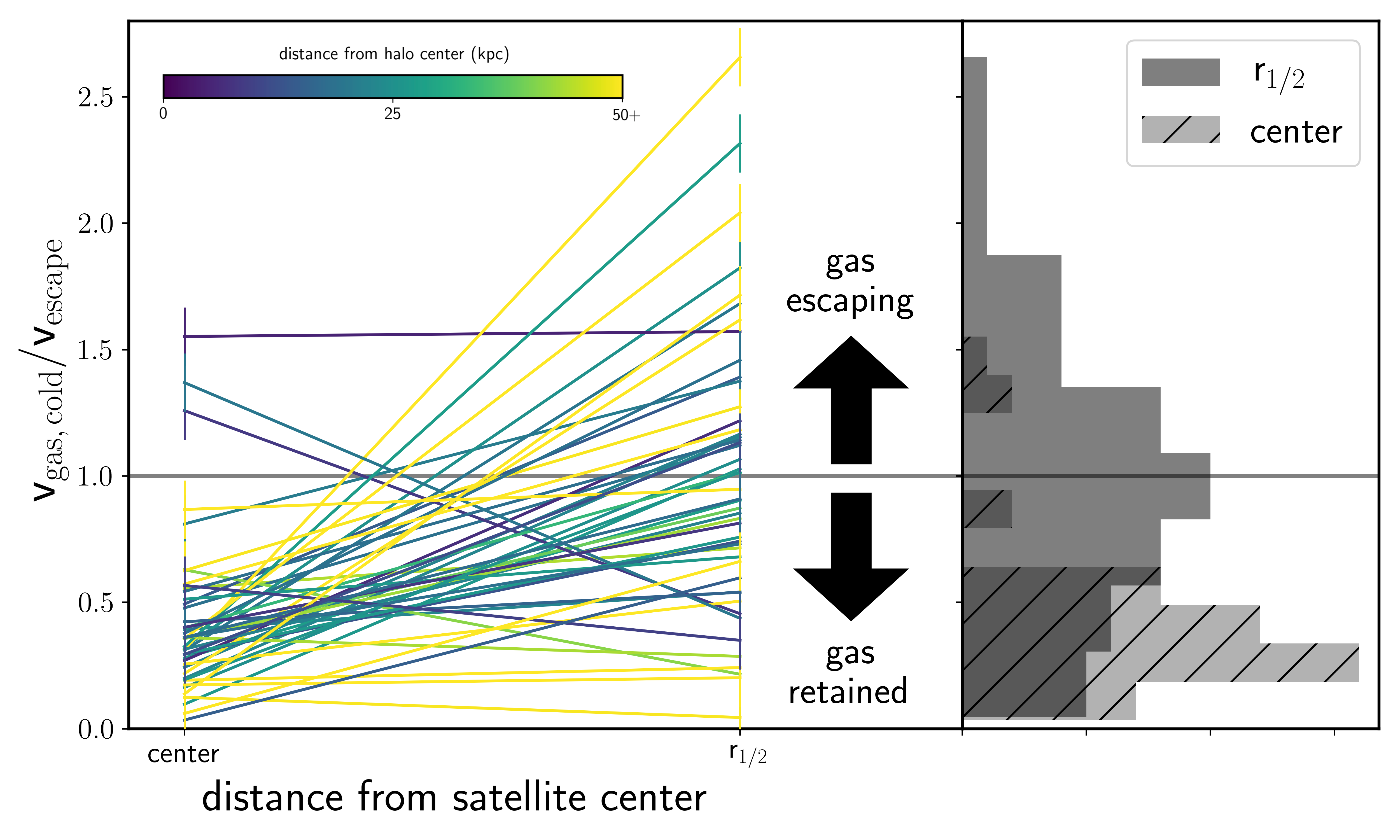}
\caption{Left: The velocity of the cold gas, in the rest-frame of the satellite divided by the velocity needed to escape the gravitational potential of the satellite is shown for the \numberofsatellites~satellite galaxies at $z = 2$. This quantity is shown at the center of the satellite and at 1 half-mass radius ($r_{1/2}$). The rest-frame of the satellite is defined by the bulk stellar velocity. At face-value, gas is removed where v$_{\rm{gas, cold}}$/v$_{\rm{escape}}$ $\ge$ 1 (grey line). The lines are color-coded by the distance of the satellite from the central, as in Figure \ref{fig:cold_stellar_mass}. Right: The distribution of v$_{\rm{gas, cold}}$/v$_{\rm{escape}}$ at the center and r$_{1/2}$ of the satellites is shown. \label{fig:vescp}}
\end{figure*}

\begin{figure*}
\includegraphics[width=\textwidth]{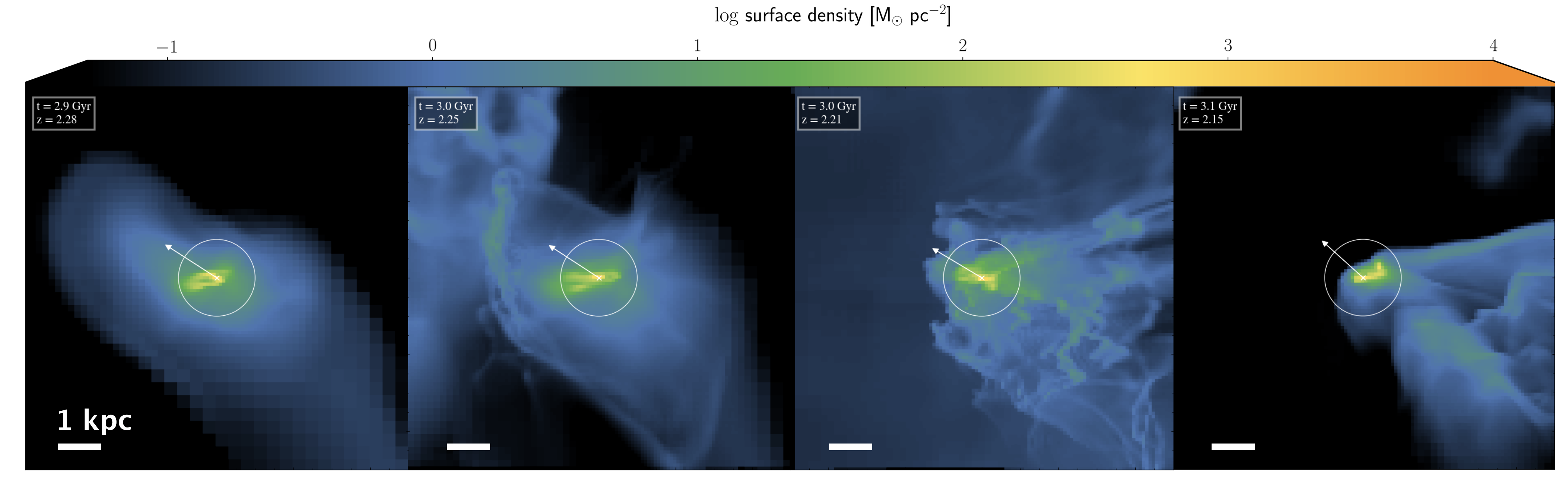}
\caption{A satellite galaxy interacts with a sheet in the circumgalactic medium of the Cyclone halo. Ram pressure stripping removes a large fraction of the gas in the outskirts of this galaxy over a relatively short period of time. This example highlights the strong, rapid, and stochastic nature of ram pressure in the FOGGIE simulations. The white bar marks 1 kpc and the white arrow shows the direction of motion of the satellite. The frames are separated by 50 Myr. \label{fig:ram_pressure_example}}
\end{figure*}

\subsubsection{Tracking Satellites Back in Time}

We track the positions and motions of the satellite galaxies identified in the previous subsection to earlier times, at $z>2$.

For each satellite, we select the one thousand oldest star particles within 1 kpc of its center at $z=2$. These stars are hereafter referred to as ``anchor stars" and are used to backtrack the motion of the satellite through the simulation box. The sample size of one thousand is large enough to calculate reliable sample statistics---population-averaged positions and velocities. 

At each timestep and for each satellite, we determine an initial position using the center of mass of the anchor stars. We next calculate the refined center of mass and mass-weighted bulk motion of all stars within a small 0.5 kpc sphere around that initial position---including stars that are not anchor stars. We record these as the position and velocity of the satellite.

We find that this is a robust method for tracking the center of the satellite, even during passages of the satellite within 1 kpc of the central galaxy. By construction, the anchor stars are retained in the center of the satellite galaxy at the last snapshot at $z=2$. By selecting anchor stars at late times and tracing them {\emph{back}} in time instead of forward, we alleviate the complication of strong dynamical interactions removing anchor stars from the satellite---which would obfuscate the true center. If the anchor stars branch into multiple distinct clusters in the history of the simulation (i.e., as galaxies that would eventually merge to form the $z=2$ satellite), then we select the most massive of these clusters as the main branch of the satellite.

At each timestep, we measure the stellar and gas mass radial profiles of each satellite galaxy. The total cold gas mass and stellar mass of the FOGGIE satellite population is shown in Figure \ref{fig:cold_stellar_mass}. The half-mass radius (r$_{1/2}$) of the satellite is defined as the 3D radius that encloses half of its total baryonic mass.

\begin{figure}
\includegraphics[width=\columnwidth]{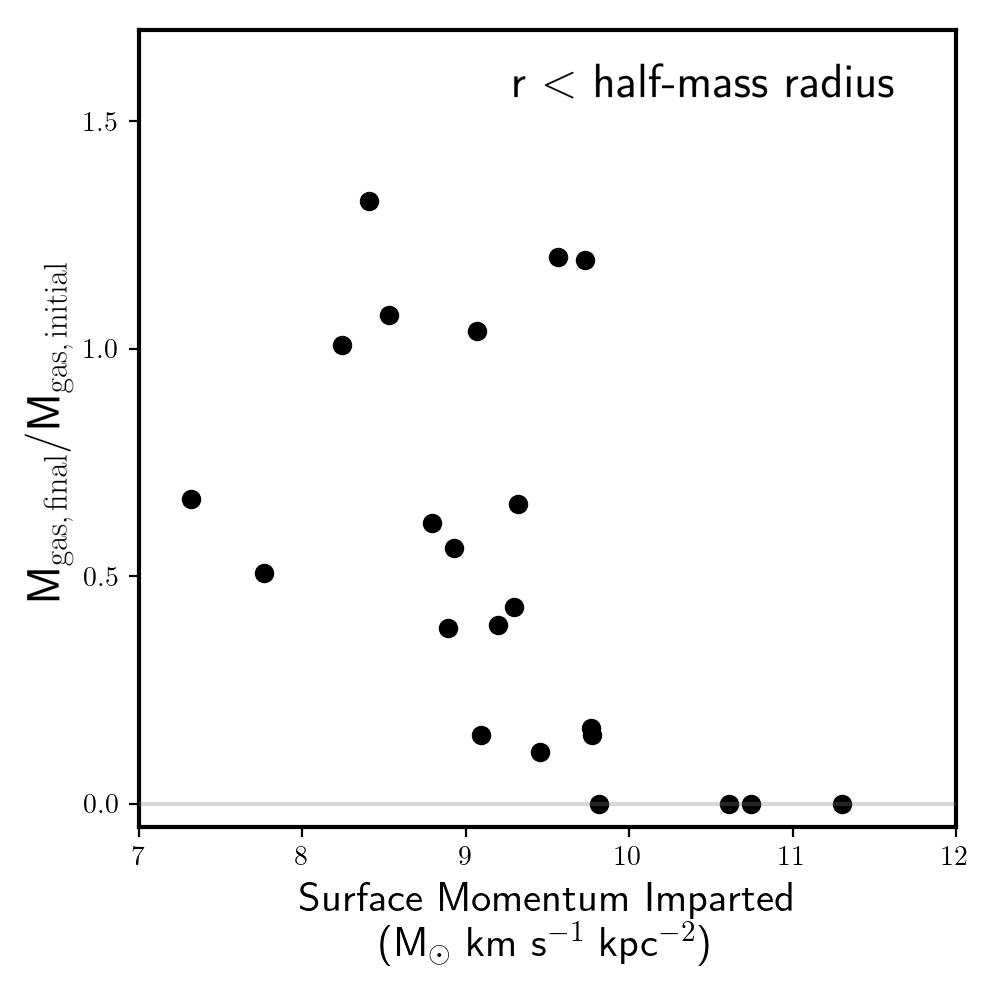}
\caption{The ratio of the final gas mass versus the initial gas mass inside one half-mass radius of the FOGGIE satellite population versus the surface momentum imparted on those satellites is shown. The simulated satellite galaxies that accumulate a higher amount of surface momentum from ram pressure lose a larger fraction of their gas. \label{fig:mom_vs_gaslost}}
\end{figure}

\subsubsection{The Accumulated Impact of Ram Pressure}
For each simulation snapshot (i.e., data outputs of the simulation separated in time by $\sim$5 Myr), we directly measure the impending ram pressure on each satellite galaxy. We use {\tt yt}\footnote{https://yt-project.org/} \citep{2011ApJS..192....9T} to create a cylinder that extends in the velocity-forward direction of the satellite and record the gas density and velocity (in the rest-frame of the satellite) of the circumgalactic medium directly ahead of the satellite galaxy. We take an average of the CGM density and velocity over a cylinder that starts at a distance of 3 half-mass radii from the satellite (to avoid confusion with the satellite's interstellar medium), extends forward by the distance the satellite will travel over the next snapshot, and is as wide as the half-mass radius of the satellite. We calculate the momentum surface density imparted through ram pressure by calculating the discrete integration in the left side of Eq. \ref{eq:stripping_crit}, where $\Delta t$ is the time interval between successive snapshots ($\sim$5 Myr). We repeat this exercise for each snapshot of each halo. 

This is a simple approach. It neglects variations in the strength of ram pressure across the face of the satellite. As the size of the satellites ($\sim$0.2 -- 0.6 kpc) are $\sim$comparable to grid resolution of the CGM, and smaller than the typical distance a satellite travels between snapshots ($\sim$1 -- 2 kpc), such variations are expected to be small and negligible compared to the time variability. Our approach also neglects ram pressure that is orthogonal to the direction of motion of the satellite (i.e., a satellite moving perpendicular to a cold stream). In general, the strength of ram pressure from the orthogonal direction is much smaller than from forward-facing direction due to lower relative velocities---the satellite's motion does not contribute to the relative orthogonal velocity.

\begin{figure*}
\includegraphics[width = \textwidth]{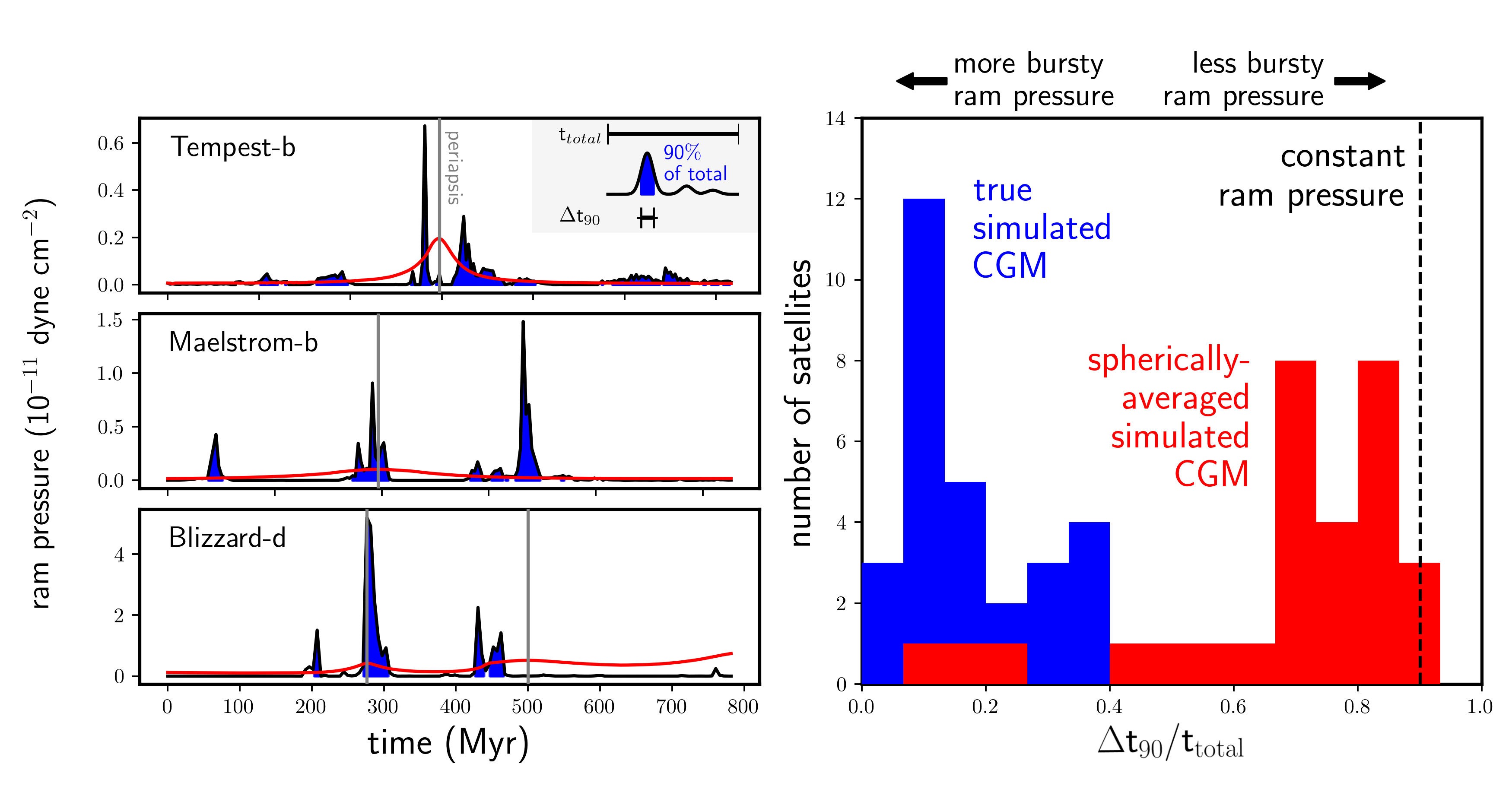}
\caption{The FOGGIE satellites experience several short, strong impulses of ram pressure as they travel through the halo of the central galaxy---on average, 90\% of the total momentum imparted by ram pressure is imparted in less than 20\% of the total time. Left: The ram pressure histories for three example satellites are shown. The true history of the satellite is shown in black (with the blue shaded region).  The blue shade indicates the shortest time needed to integrate 90$\%$ of the total momentum. The grey vertical line indicates a periapsis in the satellite's orbit. A synthetic history in which we consider a satellite traveling on the same orbit in a spherically-averaged model of the halo is shown in red. The satellites enter the high-resolution refine box at t = 0. The time integral of the curves corresponds to the momentum imparted by ram pressure. Right: The distribution of the ratio of the shortest integrated time to reach 90\% of the total (t$_{\text{90}}$) to the total time (t$_{\text{total}}$). The black dashed line indicates a ram pressure history that is constant.
\label{fig:cumulative_ram_pressure}}
\end{figure*}

\subsection{The Impact of Ram Pressure on the FOGGIE Satellites}\label{sec:rp_impact_satellites}

In Figure \ref{fig:cold_stellar_mass}, we show the cold gas mass versus stellar mass of the \numberofsatellites~satellites inside the force-refined track box at $z=2$ across all \numberofhalos simulations. Each point is color-coded by the distance of the satellite from the center of the halo. 

Satellites that are more than 50 kpc from the central galaxy all tend to be gas-rich, lying on or near the 1:1 line with stellar mass. These galaxies are on their first passage through the central host halo. Galaxies that are closer to the central show significant scatter towards high and low cold gas mass at fixed stellar mass. The stellar mass of a satellite and its location in the halo are, by themselves and/or in conjunction, insufficient bellwethers of cold gas.

In Figure \ref{fig:vescp}, we show the relative motion of the cold dense gas associated with each of the satellites at $z=2$. Specifically, we show the ratio of the velocity of the cold gas in the rest-frame of the satellite with the local satellite escape velocity. This is calculated at the center of the satellites and at one half-mass radius. As in \S\ref{sec:Ram_Pressure Stripping Toy_Satellites}, we calculate the escape velocity of the satellite assuming spherical-symmetry. Cold gas exceeding the escape velocity of the satellite is, at face value, escaping. 

At and beyond one half-mass radius, nearly half of the satellites have cold ($<1.5 \times 10^4$ K) gas velocities indicative of escape. At their centers, less than 5\% show such signatures. Interestingly, the galaxies that do show gas escaping are not preferentially found in the centers of the halos, but are instead scattered throughout the halos. 

These results indicate that ram pressure stripping is effective in both the inner and outer parts of the FOGGIE halos. However, everywhere in the halo, ram pressure is generally only strong enough to strip gas from the outskirts of the FOGGIE satellites, where the gravitational restoring pressure is lower.

In Figure \ref{fig:ram_pressure_example}, we follow a series of four snapshots of a satellite galaxy traversing through the Cyclone halo. We use this example to highlight the typical rapidity through which ram pressure impacts the FOGGIE satellites.

From beginning to end, this series spans less than 200 Myr. In the first snapshot, the satellite is passing through a relatively diffuse, kinematically-calm region of the CGM. Inside a half-mass radius, it has stellar and gas masses of 1.0 and 3.2 $\times$10$^7$ \msun, respectively. Over the next two snapshots, the satellite quickly encounters a dense filament in the circumgalactic medium with a relative speed of $\sim$330 km s$^{-1}$. Inside the filament, the ram pressure acting on the satellite is two orders of magnitude higher than it was outside. In the last snapshot, the satellite is once again passing through a part of the CGM with low ram pressure. However, it now has a gas mass of 2.3 $\times$10$^7$ \msun\,---the short encounter with the filament has removed $\sim$30\% of the satellite's gas mass. 

In Figure \ref{fig:mom_vs_gaslost}, we explore the impact of ram pressure with a larger subset of the satellites. We restrict the sample to those satellites that have been inside the refine box for at least 100 Myr---which indirectly selects those that have sampled at least $\sim10$ kpc of each halo's CGM. This timescale is comparable to the typical internal vertical period of oscillations of the FOGGIE satellite, and is the minimum time needed to allow gas to escape if imparted with a sufficient amount of momentum.

We show the ratio of the final (at $z=2$) to initial (at the time they first enter the \boxname) gas mass of the subset of satellites as a function of the integrated ram pressure (i.e., surface momentum imparted) they have accumulated. The fraction of gas mass that a galaxy retains is strongly anti-correlated with the total surface momentum imparted by ram pressure. Satellites that inherit $\gtrsim$10$^{10}$ M$_{\odot}$ km s$^{-1}$ kpc$^{-2}$ momentum per area are all gas-free.

\subsection{A Rapid Accumulation of Surface Momentum}\label{sec:rapid_accumulation}

We now examine the ram pressure histories for the full FOGGIE satellite population. In the left panel of Figure \ref{fig:cumulative_ram_pressure}, we show the ram pressure histories of three FOGGIE satellites---in the Tempest, Maelstrom, and Blizzard simulations, respectively. These histories are typical of the full satellite population and are selected to illustrate the stochastic nature of ram pressure inside the FOGGIE halos. 

The ram pressure histories are marked by several short bursts. The vertical axis is set with a linear scale to highlight the significance of these bursts. We shade the region of each history that includes 90\% of the total integral of the curve---which corresponds to the total surface momentum imparted. The short bursts are responsible for the majority of the total surface momentum that each satellite experiences throughout its trajectory.

We then consider the same satellites, on the same orbits, passing through the spherically-averaged model of their respective halos. This is more representative of what might be assumed by simple analytic or semi-analytic models of ram pressure. The spherically-averaged results are shown with red lines in the left panel of Figure \ref{fig:cumulative_ram_pressure}. By taking a spherical-average of the halo, we naturally suppress individual bursts of ram pressure (corresponding to dense and fast moving parts of the CGM that are uncharacteristic of its volume-average). As expected, the ram pressure profiles of satellites moving through the spherically-averaged medium are significantly more smooth. 

In the right panel of Figure \ref{fig:cumulative_ram_pressure}, we plot the distribution of the time it takes to {\emph{most efficiently}} (i.e., using the least amount of time) integrate to 90\% of the total surface momentum, normalized by the total time. If ram pressure were applied as a constant, we expect this distribution to peak at t$_{\rm{90}}$/t$_{\rm{total}} = 0.9$. Instead, we find that the satellite histories are distributed at much lower values---below 40\%. The median of the distribution is at t$_{\rm{90}}$/t$_{\rm{total}} = 0.18$, indicating that 90\% of the total surface momentum imparted is done so in less than 20\% of the total time, on average.

We show the same distribution of t$_{\rm{90}}$/t$_{\rm{total}} = 0.9$ for the satellites moving through the spherically-averaged model of the FOGGIE halos. The difference is dramatic. Satellites moving through the spherically-averaged model of the FOGGIE CGM accumulate momentum in a much less bursty manner. The distribution peaks near the value predicted by assuming a constant ram pressure, with a tail towards slightly lower t$_{\rm{90}}$/t$_{\rm{total}}$. Comparing the distributions from the real simulated CGM (blue) with the spherically-averaged model of the CGM (red) reveals a significant shortcoming in the spherically-averaged approximation.

To summarize, short bursts, which are stochastic in nature, account for the majority of the real cumulative impact of ram pressure. This behavior can not be captured in spherically-averaged models of the halos.

\section{Conclusions}\label{sec:conclusions}

We study ram pressure stripping in six simulations of Milky Way-like halos from the Figuring Out Gas \& Galaxies In Enzo (FOGGIE) project. We find highly stochastic ram pressure in the simulated circumgalactic medium (CGM) of these halos. The ram pressure history of a satellite galaxy passing through the CGM depends sensitively, and stochastically, on the specific path it takes. This behavior is not captured in traditional spherically-averaged models of the circumgalactic medium.

The simulations used here adopt a novel refinement technique to impose resolution at and above the gas cooling length. This allows them to efficiently sample the circumgalactic medium at high and physically-meaningful resolutions, all in a full cosmological context. In the CGM, the cooling length is resolved in $>$99\% of the volume and $>$90\% of the mass, revealing the multiphase and multiscale nature of the circumgalactic gas around Milky Way-like galaxies in unprecedented detail.

Our primary findings are  as follows:

\begin{enumerate}
\item In all six halos, the circumgalactic gas spans a large range in density and velocity---typically $\sim$\densityspan in density and \velocityspan in velocity. This translates into a \rpspan range in ram pressure. 

\item In each halo, we measure large trajectory-by-trajectory variations in the radial profiles of the ram pressure strength, due to the inhomogeneity of the density and velocity of the simulated CGM. These variations translate into a wide spread in the efficacy of ram pressure stripping in the halos. A spherically-averaged model of the CGM suppresses these variations and generally recovers a biased (high) inference on the strength of ram pressure and its integrated impact on satellite galaxies.

\item The total surface momentum imparted to a hypothetical test particle on a radial orbit through these halos is large enough to strip the outskirts (beyond 1 effective radius) in moderately massive satellites ($<10^{7} -10^{8}$  \msun) in a majority of trajectories and in massive satellites ($<10^8-10^{9}$ \msun) in a small percentage of trajectories.

\item All six halos are characterized by short-timescale ram pressure stripping. Strong impulses account for a large fraction of the total momentum imparted onto the simulated satellites. On average, 90\% of the total surface momentum imparted onto the satellites is done so in less than 20\% of the total time.

\end{enumerate}

These results reveal an erratic mode of ram pressure stripping in Milky-Way like halos at high redshift---one that is not captured by a smooth spherically-averaged model of the circumgalactic medium. To fully understand and model the origin and evolution of satellite populations around galaxies like the Milky Way and their galaxy groups, one must consider the multiscale, multiphase and variable nature of its circumgalactic gas.

\section*{Acknowledgements}
This study was primarily funded by the National Science Foundation via NSF AST-1517908. 
RCS appreciates support from a Giacconi Fellowship at the Space Telescope Science Institute, which is operated by the Association of Universities for
Research in Astronomy, Inc., under NASA contract NAS 5-26555.
BWO was supported in part by NSF grants PHY-1430152, AST-1514700, and OAC-1835213, by NASA grants NNX12AC98G, NNX15AP39G, and 80NSSC18K1105, and by HST AR \#14315. 
RA and CL were supported by NASA grants 80NSSC18K1105 and HST GO \# 15075. 
YZ acknowledges support from the Miller Institute for Basic Research in Science at the University of California, Berkeley.
Resources supporting this work were provided by the NASA High-End Computing (HEC) Program through the NASA Advanced Supercomputing (NAS) Division at Ames Research Center and were sponsored by NASA's Science Mission Directorate; we are grateful for the superb user-support provided by NAS. Resources were also provided by the Blue Waters sustained-petascale computing project, which is supported by the NSF (award number ACI-1238993 and ACI-1514580) and the state of Illinois. Blue Waters is a joint effort of the University of Illinois at Urbana-Champaign and its NCSA. This research made use of Astropy\footnote{http://www.astropy.org}, a community-developed core Python package for Astronomy \citep{astropy:2013, astropy:2018}. This research made use of Photutils, an Astropy package for
detection and photometry of astronomical sources \citep{larry_bradley_2019_3568287}. This work benefited from the butterfly emoji on Slack. 
Computations described in this work were performed using the publicly-available Enzo code, which is the product of a collaborative effort of many independent scientists from numerous institutions around the world.

\bibliography{FOGGIE_rampressure.bib}{}

\end{document}